\documentstyle[12pt]{article}
\begin{document}
\hskip 10truecm { $ \scriptstyle DSF-93/1 $ }

\hskip 10truecm { $ \scriptstyle INFNNA-IV-93/1 $ }
\vskip 1.0truecm
\centerline{\bf A MASS SPECTRUM FOR LEPTONS AND QUARKS}
\vskip 1cm
\centerline{Gianpiero Mangano$^{(a,b)}$ and Francesco Sannino$^{(a)}$}
\vskip 1cm
{}~~~~~{\it (a) Dipartimento di Scienze Fisiche, Universit\'a di Napoli, Mostra
d'Oltremare}

{}~~~~~~{\it Pad.19, I-80125, Napoli, Italy.}
\vskip 1cm
{}~~~~{\it (b) INFN, Sezione di Napoli, Mostra d'Oltremare Pad.16,
 I-80125, Napoli, Italy.}
\vskip 3cm
\centerline{\bf ABSTRACT}
\vskip 1cm
A suggestive phenomenological relation is found among the lepton and
quark masses
which suggests a new insight into the problem of generations
and also gives a prediction for the top quark mass in agreement with the
range of values which is actually expected.
\vskip 3cm

\vfill\eject

A complete and satisfactory theoretical description of the observed mass
 spectrum of {\it elementary particles}, quarks and leptons, is still
probably one of the most challenging problem in particle physics.
As well known in the electroweak standard model of Glashow, Salam and
Weinberg (GSW) all masses of fermions and of gauge bosons $W^{\pm}$ and
$Z^0$ are produced, through the Higgs-Kibble mechanism, by a symmetry
breaking phenomenon and a corresponding vacuum expectation value $v$ of
the Higgs scalar field, with $v=(\sqrt{2}G_F)^{-{1\over 2}}\sim 246~~GeV$.
However, since the Higgs-fermions Yukawa couplings are free parameters of the
model, the fermion masses are arbitrary and still in the realm of the
experiments. A legitimate point of view is therefore to think that the GSW
model is only a low energy effective version of a more fundamental theory and
the {\it deus ex machina} Higgs boson should be reinterpreted as a composite
state or completely disappear from the low-energy theory.

In this direction
goes the recent $t\overline{t}$ condensate model [1]: although
the prediction for the top quark mass seems too high, $m_t\geq 250~~GeV$,
in view of the recent $Z$ boson data from LEP, this model strongly suggests the
 emergence of a symmetry breaking \`a la Nambu Jona-Lasinio (NJL)[2] as more
fundamental.
It has been found for example that a NJL model reproduces all the features of
the GSW model as an effective theory and also predicts the following sum rules
for quark and lepton masses [3]:
$$\left[{{\sum_f m_f^2}\over {8N_g}}\right]^{1\over 2}={{m_W}\over {\sqrt{3}}}
\eqno (1)$$
$$\left[{{\sum_f m_f^4}\over {\sum_f m_f^2}}\right]^{1\over 2}=
{{m_H}\over {2}}
\eqno (2)$$
with $N_g$ the number of generations and the sum is over the $8\cdot N_g$
massive fermions.

A different, very interesting scenario seems to come out by formulating the
standard model on a Planck lattice [4]: it has been shown in this case that, by
adding NJL quadrilinear terms to the
simple lattice transcription of the GSW lagrangian, a kind of $t\overline{t}$
condensate model is produced but the low energy scalar particle (the Higgs)
has a mass pumped up to the Planck scale so it disappears from the low energy
theory.

Strictly related to the problem of fermion mass spectrum is of course
 the following very naive question: why three fermion generations?
In the standard model again there is no answer to this interrogative: the index
of generation just indicates the number of copies of the theory at
different energy scales
made by a Xerox machine. Moreover the elements of
 the Cabibbo-Kobayashi-Maskawa matrix, which mix the generations in weak
 charged currents, are unpredicted.

There are two possible attitudes towards
this problem: the first one is to assume that the existence of many
generations means that the original gauge group of symmetry is larger than
$SU(3)_c\times SU(2)_L\times U(1)_Y$, larger enough to include also the
generation degree of freedom as a charge, like colour, weak isospin and weak
hypercharge, and that this symmetry undergoes a {\it spontaneous} break
(for a brief survey see for example [5] and references therein).
The price for this of course is the proliferation of gauge bosons and Higgs
particles necessary (up to now!) to the symmetry breaking.

\noindent
The second one, which is also our attitude, would ascribe to the still unknown
dynamical mechanism which generates fermionic masses the
capability of predicting the observed spectrum. In a recent paper [6] some
empirical relations among the square root of lepton and quark masses have been
proposed:
$${{\sqrt{m_c}}\over {\sqrt{m_u}}}-{{\sqrt{m_s}}\over {\sqrt{m_d}}}=
{{\sqrt{m_\mu}}\over {\sqrt{m_e}}}\eqno(3)$$
$${{\sqrt{m_t}}\over {\sqrt{m_c}}}-{{\sqrt{m_b}}\over {\sqrt{m_s}}}=
{{\sqrt{m_\tau}}\over {\sqrt{m_\mu}}}\eqno(4)$$

\noindent
the second one being quite well satisfied with a mass for the top quark
$m_t\sim 131~GeV$. It has been also stressed that the simplicity and regularity
 of these relations may indicate that quarks and leptons of the second and
third generation are excited states of the corresponding ones in the first
generation. However (3) and (4) only represent two {\it sum rules} which the
fermion masses should obey for certain unknown
properties.

\noindent
The actual situation is very similar to the one relative to atomic spectra
just before the famous paper of Bohr in 1913: in that case, as well known, the
radiation frequencies corresponding to atomic transitions were empirically
found
 to be expressible in terms of {\it natural numbers} and of only one constant,
the Rydberg constant, for each kind of atomic species. With this historical
 reference in mind, we believe that also a purely empirical but {\it appealing}
relation which fits well the quark and lepton masses could be very useful for
our understanding of mass generation mechanism, in particular if this relation
is:
\vskip .5cm
\noindent
{}~~~~~1) very simple
\vskip .5cm
\noindent
{}~~~~~2)expresses masses only in terms of natural numbers and of a constant.
\vskip .5cm
\noindent
This is exactly the aim of this letter.

Let us first consider the charged
leptons: from the Review of Particle Properties '92 [7]:
$$m_e=(51099906\pm 15)\cdot 10^{-8}~MeV$$
$$m_\mu=(105658389\pm 34)\cdot 10^{-6}~MeV$$
$$m_\tau=(1784.1 ^{+2.7}_{-3.6})~MeV$$
\noindent
These masses are quite well fitted by the following relations:
$$k_E\cdot ln(m_\mu/m_e)=ln(2)~~~~~~~~~~~~~~~~~~~~
k_E \cdot ln(m_\tau/m_e)=ln(3)\eqno(5)$$

\noindent
The value of the $k_E$ parameter is only slightly different in the two cases:
$k_E\sim 0.130$ for the muon and $k_E\sim 0.135$ for the $\tau$. The
relations (5)
are of course meaningful only if $k_E$ has the dignity of a {\it constant},
equal
for both $\mu$ and $\tau$: the mentioned
quite small difference ($\sim 4\%$) may be
however naively understood in terms of radiative correction effects. We will
come back on this point very soon asking what could be the meaning of the $k_E$
parameter. Equation (5) may be
written in the following form:
$$m_p=m_1~p^{1/k_E}~~~~~~~~~~~~~~~p=1,2,3 \eqno(6)$$

\noindent
About this {\it spectral} formula we would make two comments:
\vskip .5cm
\noindent
{}~~~~~1) We are not saying anything about the mass $m_1=m_e$ which represents
the fundamental state.
\vskip .5cm
\noindent
{}~~~~~2) The particular simple form of the spectrum (6) is very reminiscent of
 some non perturbative dynamical mass generation if $k_E$ plays the role or is
a simple function of the coupling constant of a new kind of effective
interaction. To be explicit it seems quite natural to us the comparison
 of (6) with
the well known expression of the gap parameter $\Delta$ in BCS theory for
superconductivity:
$$\Delta\sim \omega_D e^{-1/g}$$

\noindent
with $\omega_D$ the Debye frequency and $g$ the average interaction energy of
 an electron interacting with unit energy shell of electrons on the Fermi
surface.

\vskip .5cm

We would now extend our consideration to quark masses: there are in this
 case two kind of difficulties. The first one is that the top quark mass is
still unknown, though it is expected to be in the range $150^{+23}_{-26}
\pm 16~GeV$ [7]: for
the up quarks therefore we only have at hand the two masses of u and c.
However, as we will see,
this circumstance will allow us to test our spectrum by
giving a value for $m_t$ in the mentioned range.

\noindent
The second problem is due to the fact that the masses of the lighter quarks u,
d
 and s are only grossly known: from chiral perturbation theory one may extract
an estimate of the so called current masses [7](for a review see [8]):
$$m_u=5\pm 3~MeV$$
$$m_d=10\pm 5~MeV\eqno(7)$$
$$m_s=200\pm 100~MeV$$

These values are obtained by considering the effect of strong interactions at
low energy ($\leq 1~~GeV$): in other terms it is assumed that the dynamic
corresponding to chiral symmetry breaking in strong interaction is responsible
for the mass spectrum of light quarks. In this scheme however the heavy quark
masses cannot be similarly treated so they remain substantially as free
parameters.
It is therefore legitimate to wonder what would happen to quark
masses if we turn off strong interactions. In this case, apart from the
electric charge, quarks are similar to leptons, so it would be quite natural
to assume that

\vskip .5cm
{}~~~1) quarks generations should satisfy a mass spectrum similar to the
lepton one

\vskip .5cm
{}~~~2) the two fundamental states $m_u$ and $m_d$, should be of the order
of magnitude of $m_e$.

\vskip .5cm

These hypothesis are based on the idea that the mechanism responsible for mass
generation is {\it universal}. We would stress however, once again, that, by
assuming 2), we are dealing with the {\it bare} masses for the low-lying
states u and d. In other words we are neglecting
all {\it dressing} effects at low energy due
to QCD, which would strongly affect the values of $m_u$ and $m_d$.

For d-like quarks (d,s,b) a spectral relation similar to (5) holds:
$$k_D\cdot ln(m_s/m_d)=ln(2)~~~~~~~~~~~~~~~~~~~~
k_D \cdot ln(m_b/m_d)=ln(3)\eqno(8)$$

\noindent
With $m_d=m_e$, $m_s=200~MeV$ and $m_b=5~GeV$ the values of $k_D$ in the
relations (8) are
again quite equal, $k_D\sim 0.116$ for the s and $k_D\sim 0.119$ for the b,
showing, as for charged leptons, an intriguing regularity.
Taking $m_c=1.5~GeV$ and assuming that also for u-like quarks a similar
relation holds:
$$k_U \cdot ln(m_c/m_u)=ln(2)\eqno(9)$$
we find $k_U\sim 0.087$. Making use of this value, from:
$$k_U \cdot ln(m_t/m_u)=ln(3)\eqno(10)$$
we find a value for $m_t$ of the order of $m_t\sim 150~GeV$. This estimate is
very rough and only indicative due substantially to the fact that though also
in this case we may expect only a slightly different value of $k_U$ at the
top mass scale, however the resulting value of $m_t$ may change strongly due
to the exponentially dependence on this parameter.

\noindent
It should be stressed however that, by assuming (10), the right order of
magnitude for $m_t$ has been obtained.

A legitimate question arises at
this point: what is the role of the adimensional constants $k_E,~k_U$ and
$k_D$?
It is quite immediate, by remembering what we said at the beginning, to figure
these parameters as the equivalent of the Rydberg constant in the atomic
spectra; this analogy, if valid, has two immediate consequences:
\vskip .5cm
\noindent
{}~~~~~~1) like the Rydberg constant
$\displaystyle{R={{m_{e} e^4}\over {2{\hbar}^2}}}$ the
$k$'s should be a function of the coupling constant of the interaction
responsible for fermion mass generation, containing in this way informations
 on the dynamics related to this interaction.
\vskip .5cm
\noindent
{}~~~~~~2) in this hypothesis, as already said, the slight change in the values
 of both $k_E$ and $k_D$ in passing from the second to the third generation
may be easily understood as  a consequence of the running character of the
coupling constant $g$ due to quantum effects.
\vskip .5cm
\noindent
To test the consistency of this assumption we have considered some simple
dependence of $k$ on $g$, namely $k\sim const \cdot g$, $k \sim const\cdot g^2$
and $k \sim const \cdot g^4$ and a standard scaling behaviour for the coupling
 constant at one loop:
$${1 \over {g^2( M)}}-{1\over {g^2(m)}}={{\beta} \over {8 \pi^2}}
ln \left({M\over m}\right)\eqno(11)$$

\noindent
{}From the lepton data the value of $\beta$ can be deduced and then applied
to the down quark spectrum to calculate, from the value of $k_D$ at
the bottom mass scale and using (11), the value for the strange quark mass:
similarly from the
value of $k_U$ at the c-quark mass it is possible to deduce the expected value
 for the top quark mass. The results of this simple procedure are illustrated
in
Table 1 for the three different cases under examination: the errors quoted on
$\beta$, $m_s$ and $m_t$ are respectively due to the $m_\mu$ and $m_\tau$,
$m_b$, and $m_c$ experimental errors. The results obtained for $m_s\sim
200~MeV$
show the consistency of the hypothesis of a unique dynamical mechanism which
lies under the mass generation of leptons and quarks, i.e. a unique
$\beta$: moreover the values for $m_t$ are very reasonable and of course in the
right order of magnitude.

\noindent
All we have seen so far still leaves open two questions which could be related
each other, as we will see. First of all it could seem unsatisfactory that
three different constant $k_E$, $k_U$ and
$k_D$ without any explicit relationship among them are necessary to account
for the observed spectra of, respectively, charged
leptons, u-like and d-like quarks. The second point we would stress is that up
to now we did non face at all the problem of neutrino masses. This latter
question is still an open one. It is possible that the neutrinos could be
massless for some theoretical reason, for example chiral invariance:
in other frameworks,
in particular in the grandunified models, also motivated from cosmological
reason, there is place for Dirac and
Majorana mass terms.

{}From an experimental point of view all
we have are some upper limits on the masses:
$$m_{\nu{_e}} < 7.3~~eV$$
$$m_{\nu_{\mu}} < 0.27~~MeV\eqno(12)$$
$$m_{\nu_{\tau}} < 35~~MeV$$

\noindent
Let us now come back to our spectral relations: is it possible to apply similar
considerations to neutrino masses ?

\noindent
We would begin by noticing the existence of
a sort of {\it unitarity} relation: if we calculate the values of the $k$
 parameters at the fundamental scale $m_e$ by using (11) one gets the
following  relation grossly satisfied:
$$k_E^2\sim k_D^2+k_U^2\eqno(13)$$

\noindent
In the three cases considered in Table 1 the difference between the left
and right-hand side in (13) is of the order of $4 \cdot 10^{-3}$, of the
order however of $k_U^2$, {\it the right-hand side being always larger}.

\noindent
The sense of relation (13), very appealing for the
{\it universality} character which recovers to the $k$ parameters,
if meaningful, may be twofold:
\vskip .5cm
\noindent
{}~~~~~~1) one possibility is that (13) has to be assumed {\it exact},
although this is not verified in our naive calculations. The meaning of
this assumption is quite clear: it is to say that the
parameters $k_E$, $k_D$ and $k_U$ are not equal because a rotation in
the $k$ plane is necessary to get the quark mass eigenstates. In this
scheme there is no natural and simple way to account for neutrino masses.
\vskip .5cm
\noindent
{}~~~~~~2) The other possibility is that the small difference between the two
members of (13) is due to the presence of a contribution $k_N^2$ coming
from neutrinos: in this sense a similar rotation to the one which holds
for quarks should happen also in the lepton sector and non-zero neutrino
masses should be expected, again with a spectrum similar to (6).
\vskip .5cm

Let us consider first this latter case. In Table 1 we have reported
the values $k_N$ which are necessary to get an exact unitarity relation:
$$k_E^2+k_N^2= k_D^2+k_U^2\eqno(14)$$

\noindent
By using the machinery given by (11) and assuming for neutrinos a spectrum:
$$m_{\nu_p}=m_{\nu_1} p^{1/k_N}~~~~~~~~~~~~~~~~~~~p=1,2,3\eqno(15)$$
saturating moreover the $\nu_\tau$ mass upper bound one gets the following
extreme limits (the case is $k\sim g$):
$$m_{\nu_\tau}< 35~~MeV$$
$$m_{\nu_\mu}< 80~~keV\eqno(16)$$
$$m_{\nu_e}< 2~~eV$$

\noindent
The same order of magnitude are found
with $k\sim g^2$ and $k\sim g^4$.

A rather different situation corresponds to case 1).
We have already said that to assume (13) as an exact relation due to the
{\it universal} character of fermion mass generation may naively correspond to
assume massless neutrinos: it has been proposed that this
circumstance could be related to
their neutrality [9]. In this case, which would relate the phenomenon of mass
generation to the electric charge of the elementary particles, one should also
expect that the lower states $m_e,~m_d,~$ and $m_u$ should be different.
We have implemented this appealing idea in its simpler form, namely we have
considered that the fundamental states e, d and u have a mass simply
proportional to corresponding value of electric charge:
$$m_u={2\over 3}m_e~~~~~~~~~~~~m_d={1\over 3}m_e\eqno(17)$$
Applying to this case all considerations discussed so far we get the results
summarized in Table 2. Notice that a lower
value for $m_s$ is obtained $m_s \sim 140~MeV$ while an heavier top quark is
predicted. Although also in this case the unitarity relation (13) is only
approximately satisfied it should be stressed that the situation is
better than in the case $m_u=m_d=m_e$: in particular in the case $k\sim g^4$
the value for $k_N$ is compatible with zero.

We are now at the conclusions. We have shown a reasonable spectral relation
which fits well lepton and quark masses if a mass for the d and u quarks
of the order of $m_e$ is assumed. The consistency of this relation seems
encouraged by a prediction for the top quark mass in the expected range. We
do not have at hand so far any dynamical picture of how such a spectrum could
come out: this  point seems of course particularly stimulating. It is not
still clear the role of the mentioned unitarity relation in the form (13) or
(14): the universality character of the underlying interaction which generates
the fermion masses which emerges from this relation could in principle mean
that a finite mass for neutrinos is expected (see(14)). What we have shown is
that, by using the idea of universality among the k's parameters and the
$\nu_\tau$ mass experimental limit, assuming the validity of our spectral
formula also for neutrinos, reasonable upper limits may be deduced on the
electron and muon neutrino masses.

We have however also
considered the fascinating but at the moment very speculative
case of a charge-mass relationship: in this
scenario one would reasonably predict massless neutrinos and would
expect different
values for the masses of the fundamental states $m_e$, $m_u$ and $m_d$. In a
simple assumption of a direct proportionality to the electric charge,
it comes out that the (13), although not exact, is quite well satisfied.

We believe that the consideration exposed in this paper may lead
to a new insight in the fascinating problem of fermion masses and
generations.

\vfill\eject
\centerline{\bf Acknowledgments}
\vskip 1cm
It is a pleasure to thank Prof. F. Buccella, to whom we dedicate this paper,
for his valuable help and criticism.
We would also thank G. Miele, L. Rosa and P. Santorelli for useful discussions.

\vskip 2cm
\centerline{\bf References}

\vskip 1cm
\noindent
[1] Y. Nambu, in "1988 International Workshop on New Trends in Strongly
Coupled Gauge Theories", editors M. Bando et al., World Scientific, Singapore
1989; W. Bardeen, C. Hill and M. Lindner, Phys. Rev. {\bf D41} (1990) 1647.
\vskip .5cm
\noindent
[2] Y. Nambu and G. Jona-Lasinio, Phys. Rev. {\bf 122} (1961) 345.
\vskip .5cm
\noindent
[3] H. Terazawa, Y. Chikashige and K. Akama, Phys. Rev. {\bf D15} (1977) 480.
\vskip .5cm
\noindent
[4] G. Preparata and S.S. Xue, Nucl. Phys. {\bf B}, Proceedings Supplement for
"Lattice '91" (1992); Preprint MITH 92/07.
\vskip .5cm
\noindent
[5] L. O'Raifeartaigh, "Group Structure of Gauge Theories",
Cambridge University
Press, 1986.
\vskip .5cm
\noindent
[6] H. Terazawa, Mod. Phys. Lett. {\bf A7} (1992) 1879.
\vskip .5cm
\noindent
[7] Review of Particle Properties '92, Phys. Rev. {\bf D45} (1992).
\vskip .5cm
\noindent
[8] J. Gasser and H. Leutwyler, Phys. Rep. {\bf 87} (1982) 77.
\vskip .5cm
\noindent
[9] K. Akama, Prog. Theor. Phys {\bf 84} (1990) 1212.

\vfill \eject
\centerline{\bf Table Captions}
\vskip 1cm
$\bullet$ {\bf Table 1}:
The values of $m_s$, $m_t$ and $k_N(m_e)$ in the case $m_u=m_d=m_e$
\vskip 1cm
$\bullet$ {\bf Table 2}:
The values of $m_s$, $m_t$ and $k_N(m_e)$ in the case $m_u={2\over3}m_e$,
$m_d={1\over 3}m_e$

\newpage

$~$
\vskip 2cm

\centerline{\bf Table 1}
\[ \begin{array}{|c|c|c|c|c|} \hline
&&&&\\
k(g)&\beta &m_s&m_t&k_N (m_e) \\
&&(MeV)&(GeV)&\\
&&&&\\
\hline
&&&&\\
\sim g&-8.94\pm0.07&203\pm8&120\pm26&0.063\pm0.003\\
&&&&\\
\hline
&&&&\\
\sim g^2&-0.592\pm0.005&208\pm9&100\pm22&0.059\pm0.003\\
&&&&\\
\hline
&&&&\\
\sim g^4&-0.1076\pm0.0008&210\pm9&90\pm20&0.056\pm0.003\\
&&&&\\
\hline
\end{array}\]

\newpage
\centerline{\bf Table 2}
\[ \begin{array}{|c|c|c|c|c|} \hline
&&&&\\
k(g)&\beta &m_s&m_t&k_N (m_e) \\
&&(MeV)&(GeV)&\\
&&&&\\
\hline
&&&&\\
\sim g&-8.94\pm0.07&133\pm5&150\pm34&0.033\pm0.005\\
&&&&\\
\hline
&&&&\\
\sim g^2&-0.592\pm0.005&138\pm6&126\pm28&0.022\pm0.007\\
&&&&\\
\hline
&&&&\\
\sim g^4&-0.1076\pm0.0008&142\pm6&112\pm25&0.009\pm0.018\\
&&&&\\
\hline
\end{array}\]

\end{document}